\documentclass{aa}

\usepackage{txfonts}
\usepackage{graphicx, psfrag}
\usepackage{natbib}
\usepackage{aalongtable}

\newcommand \sg {$\sigma$~Gem}
\newcommand \chandra {{\it Chandra}}
\newcommand \cxc {{\it Chandra}}
\newcommand \xmm {{\it XMM-Newton}}

\begin{document}

\title{On Temperature and Abundance Effects During an X-Ray Flare on $\sigma$~Geminorum} 

\author{Raanan Nordon \inst{1} \and Ehud Behar \inst{2} \and Manuel G\"udel \inst{3}}

\institute{Department of Physics, Technion, Haifa 32000, Israel; nordon@physics.technion.ac.il 
      \and Department of Physics, Technion, Haifa 32000, Israel; behar@physics.technion.ac.il
      \and Paul Scherrer Institut, W\"urenlingen \& Villigen, 5232 Villigen PSI, Switzerland; guedel@astro.phys.ethz.ch
}

\date{Recived / Accepted}

\abstract{
We compare quiescent and flare X-ray spectra of the RS CVn binary \sg\ obtained with the \chandra\ and \xmm\ grating spectrometers. We find that in addition to an overall 25\%\ flux increase, which can be ascribed to variations in the system's quiescence activity over the 15 months that passed between the observations, there is a hot plasma component of $kT_e \ga$~3~keV that arises with the flare. The hot component is manifested primarily by emission from high charge states of Fe and by a vast continuum. 
The cooler ($kT_e$ $\la 2$~keV) plasma remains undisturbed during the flare. We find no significant variations in the relative abundances during the flare except for a slight decrease ($<$~30\%) of O and Ne.
}

\titlerunning{X-ray Flare on $\sigma$ Gem}
\authorrunning{Nordon et al.}

\maketitle

%%%%%%%%%%%%%%%%%%%%%%%%
\section{Introduction}
%%%%%%%%%%%%%%%%%%%%%%%%
The interplay between steady coronal emission and coronal flares has been a subject of ongoing research for many years. X-ray line resolved spectra available with \chandra\ and \xmm\ allow now for unprecedented plasma diagnostics, one of which is elemental abundance measurements.

Measurements of relevant changes could offer important diagnostics for the heating and the plasma transport process in stellar coronae. It is unclear how exactly large flares affect the abundances in stellar coronae. Although indications for changes in the metallicity of a flaring corona  were reported early on from low-resolution devices \citep[see discussion in][]{Gudel2004}, more reliable analysis had to wait for the advent of high-sensitivity and medium-to-high resolution spectrometers. Indications for increasing metallicities were found from {\it ASCA} and {\it BeppoSAX} medium-resolution observations \citep[e.g.,][]{Mewe1997, Tsuboi1998, Favata1999}. Moreover, when analysing elemental abundances of individual elements, selective, significant enhancements of low first ionization potential (FIP) elements have been reported \citep{Gudel1999, Osten2000, Osten2002}. In contrast, in several other cases, the composition of the plasma appeared to remain unchanged during flares \citep[e.g.,][]{Maggio2000, Franciosini2001}.

A crucial reconsideration of the situation  came with the advent of high-resolution spectroscopy with {\it XMM-Newton} and {\it Chandra}, but the findings so far still lack systematic trends. While \citet{audard01} found significant enhancements of low-FIP elements in a flare on HR 1099, two flares reported by Osten et al. (2003; for $\sigma^2$ CrB) and G\"udel et al. (2004; for Proxima Centauri) showed an increase of the abundances of several elements, but no selective FIP-dependence was found.

RS~CVn binary systems are bright X-ray and EUV sources owing to rapid rotation, generating a magnetic dynamo. As such, they have been studied extensively in both bands \citep[][respectively, and references therein]{Audard2003, Sanz-Forcada2002}.
The RS~CVn $\sigma$ Geminorum (HD62044, HR 2973, HIP 37629) is particularly bright and well observed at all wavelengths. For an RS CVn, it has a rather long period of 19.6045 days \citep{Duemmler}. The primary star is a K1 III type, red giant. Little is known of the secondary as it has not been detected at any wavelength, but restrictions to its mass and the low luminosity suggest that it is most likely a late-type main sequence star of under one solar mass \citep{Duemmler}.

The \sg\ system is a luminous X-ray \citep[$\log L_X \approx 31.0 \pm 0.2$~erg~s$^{-1}$;][corrected for a distance of 37.5~pc]{Yi} and radio \citep[$\log L_R \approx$ 15.40~erg~s$^{-1}$~Hz$^{-1}$, at 6~cm wavelength;][]{Drake1989} source. While most observations found it to be a relatively steady emitter, a very large flare has been detected in December 1998 with EUVE \citep{Sanz-Forcada2002}. Another flare was detected in April 2001 both in the X-ray and in the radio in which a Neupert effect: $\frac{d}{dt}L_{X} \propto L_{Radio}$ \citep{Neupert1968} was found \citep{Gudel2002}. The common explanation for the Neupert effect is that the gyrosynchrotron radio emission during the flare is due to fast particles spiraling down the flaring loop. When this particle population hits the footprint of the loop they heat the choromspheric material, which is subsequently driven into the corona where it emits X-rays. Thus, the radio emission depends on the instantaneous number of particles running through the loop, while the slowly varying X-ray emission from the surrounding chromosphere represents the total fast-particle energy converted into heat.

The detection of the Neupert effect involved only the light curve extracted from the EPIC-pn instrument on board \xmm. In this paper we investigate the X-ray spectrum emitted during the 2001 flare and compare it to a quiescent spectrum obtained in December 1999. Our main goal is to understand the effects of the flare on the thermal and chemical structure of the X-ray plasma and to examine whether they support the chromospheric evaporation scenario.

%%%%%%%%%%%%%%%%%%%%%%%%%%%%%%%%%%%%%%%%%
\section{Observations and data analysis}
%%%%%%%%%%%%%%%%%%%%%%%%%%%%%%%%%%%%%%%%%
\subsection{Data Reduction and Light Curves}

\begin{figure}
%  \begin{center}
\vspace{0.8cm}
    \resizebox{\hsize}{!}{\includegraphics{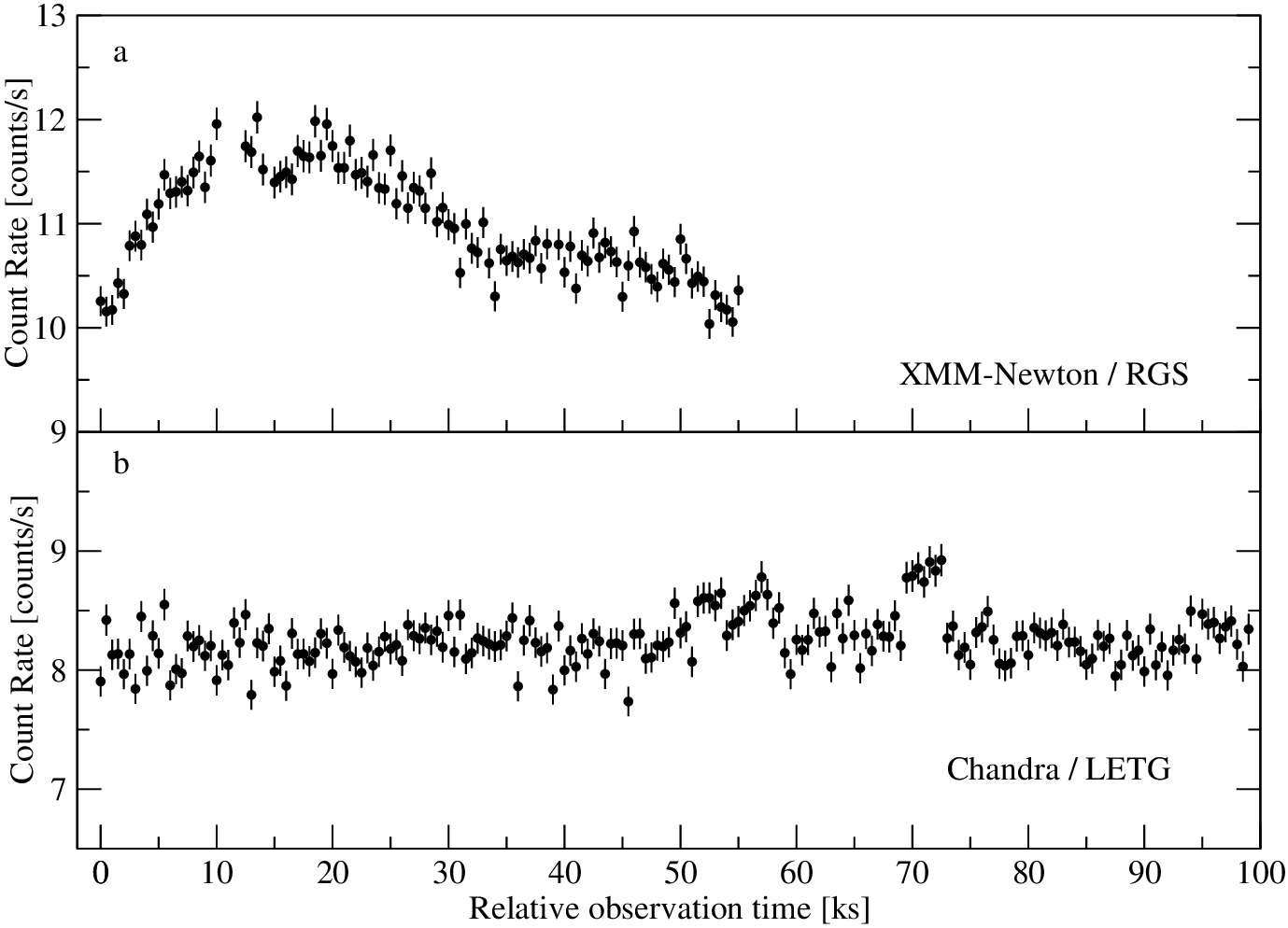}}
    \caption{Light Curves for \sg\ in time bins of 500~s. (a) April 2001 RGS (1 and 2 combined) observation (1st and 2nd orders). (b) December 1999 LETG/ACIS observation (all orders).}
    \label{fg:lightcurve}
%  \end{center}
\end{figure}

The target \sg\ was observed by \xmm\ in April 2001 for a total exposure time of 54~ks (observation start: 2001-04-06 16:24:29, end: 2001-04-07 07:54:40, UTC). The data were reduced using the Standard Analysis System (SAS) version 6.0.0.
In this analysis we use the Reflection Grating Spectrometers (RGS) in the 1st order of diffraction, which gives reliable data from 6 to 38~\AA. Line fluxes of Fe$^{24+}$ and Fe$^{25+}$ were extracted from the EPIC-pn data with the use of the XSPEC software package \citep{Arnaud1996}. Background is subtracted using off-source CCD regions. 
\chandra\ observed the target on December 1999 (observation start: 1999-12-27 18:38:39, end: 1999-12-28T23:07:07, UTC) for a duration of 100~ks with the Low Energy Transmission Grating (LETG) + Advanced CCD Imaging Spectrometer (ACIS) configuration in Continuous Clocking (CC) mode. The data were reduced using the CIAO package version 3.0.2.

The long duration flare, with an observed peak rise of 20\%, observed also by \citet{Gudel2002}, is seen in the RGS light curve presented in Figure~\ref{fg:lightcurve}. The rise in flux is roughly uniform in the entire RGS wavelength band. The flare light curve is contrasted with the flat light curve obtained from the LETG observation shown below in the same figure.

The use of CC mode is difficult for spectral extraction due to order mixing. 
Above 20~\AA, the orders become smeared and mixed and do not allow for reliable order-sorting. 
Part of the smearing could be due to cosmic ray afterglow, but since there is no background assessment for the CC mode, it cannot be subtracted in a systematic way. Therefore, only data in the range of 1.8 to 20~\AA\ are used in the present work. 
One exception is the N$^{6+}$ line at 24.78~\AA\ which can be used. Although the continuum level in this region is unreliable, the flux in the line is not affected by mixing with the 2nd order, as no major peaks occur at 12.4~\AA.
Since the target is very bright, background can be neglected.
Also, CIAO's default extraction region is too narrow for CC mode, missing out many source photons, especially in the -1 order. This is due to CC-mode gain not being well calibrated. In order to extract all of the observed source photons, a wider energy region had to be selected. In the tg\_resolve\_events task, we used: {\small OSORT\_HI=0.34 OSORT\_LO=0.20} for $m*\lambda > 7.5~\AA$ and {\small OSORT\_HI=0.08 OSORT\_LO=0.12} for $m*\lambda < 7.5~\AA$
\footnote{See the \chandra\ website at http://cxc.harvard.edu/ciao/ahelp/tg\_resolve\_events.html}. 
For the spectral analysis, only 1st order data were used. The spectra obtained from the quiescent and flare observations are shown in Figure~\ref{fg:spectra}.

\subsection{Direct comparison of flare and quiescence spectra} 

\begin{figure}
%  \begin{center}
\vspace{0.8cm}
    \resizebox{\hsize}{!}{\includegraphics{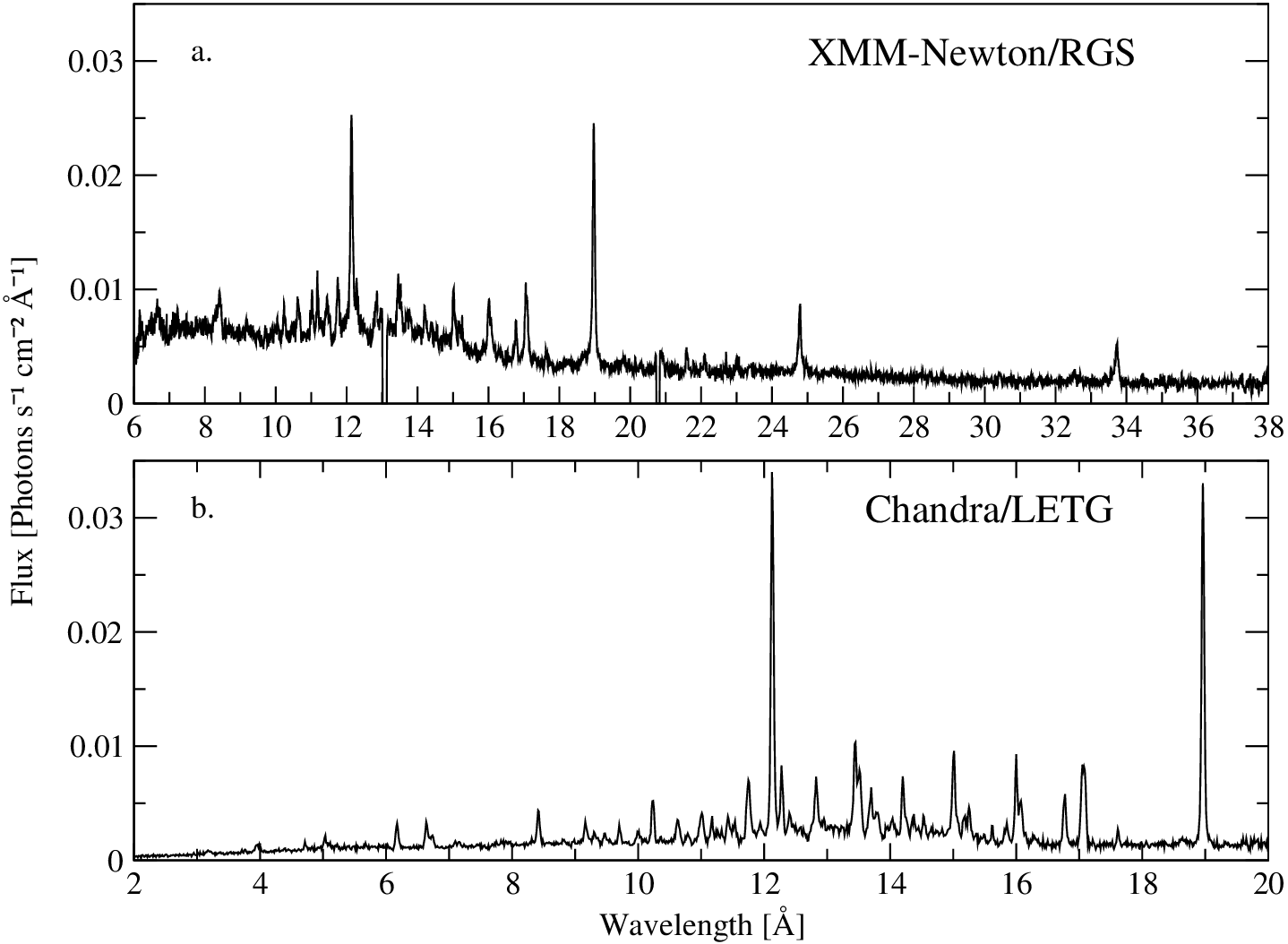}}
    \caption{\sg\ spectra. (a) Flaring \textit{XMM}-RGS spectrum, April 2001, average of the two RGS instruments in 1st order. (b) Quiescent state \textit{Chandra}-LETG spectrum, December 1999, average of the two 1st orders.}
    \label{fg:spectra}
%  \end{center}
\end{figure}

\begin{figure*}
%  \begin{center}
    \includegraphics[width=17cm]{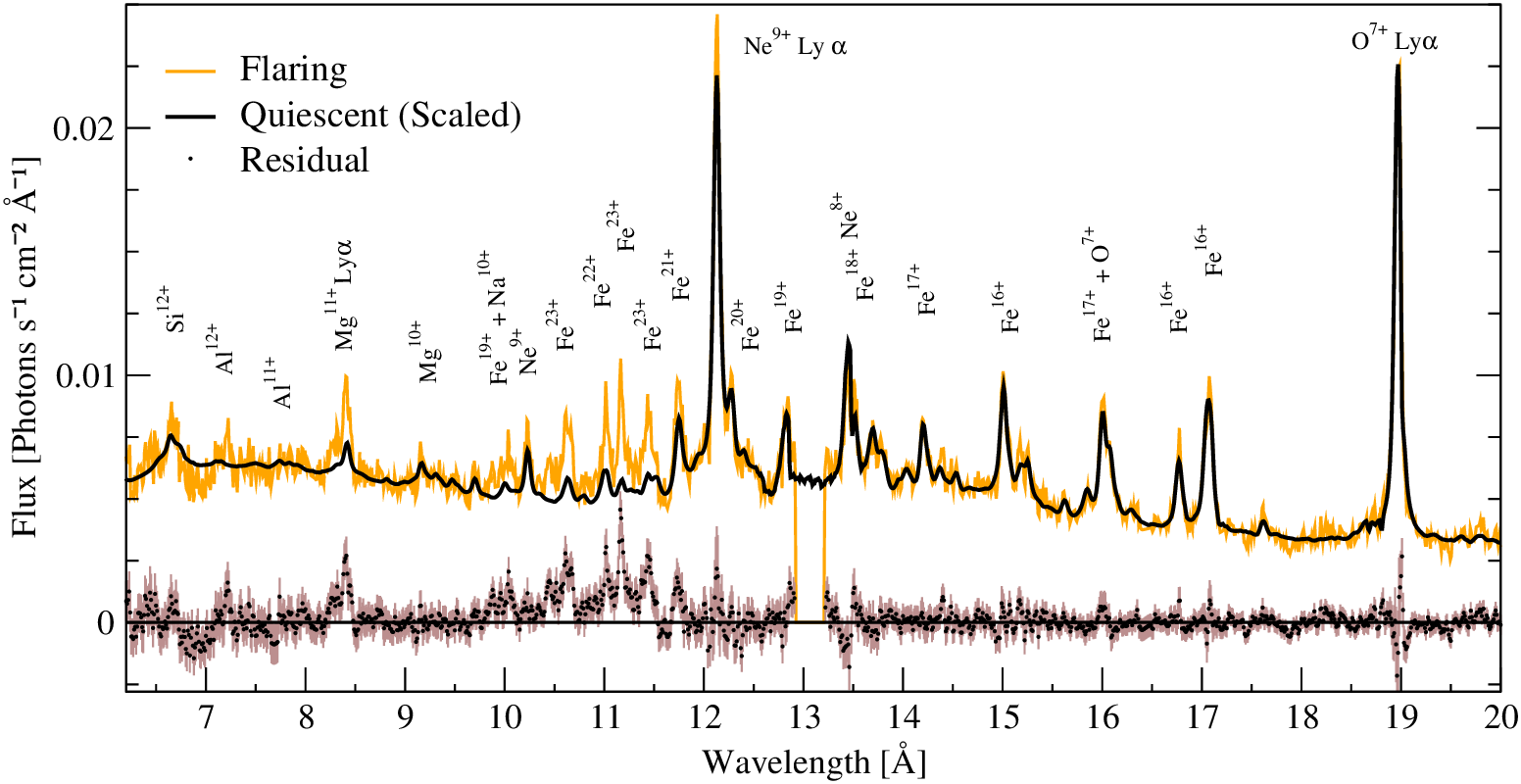}
%    \resizebox{\hsize}{!}{\includegraphics{4018f4.eps}}
    \caption{Line intesities of the flaring and quiescent states compared. The quiescent lines are multiplied by a factor of 1.25 and an ad-hoc continuum is added to match the flaring continuum. The plotted residual is the quiescent curve subtracted from the flaring spectrum, with corresponding error bars.}
    \label{fg:model_diff}
%  \end{center}
\end{figure*}

The most obvious difference between the two spectra in Figure~\ref{fg:spectra} is the vast continuum emission in the flare. While in quiescence the continuum starts to drop below 12~\AA\ , during the flare the continuum is still rising at 6~\AA\ , which is the limit of the RGS. 
Using the data from EPIC-pn and assuming the continuum is thermal bremsstrahlung, we can determine the dominant plasma component contributing to the flare continuum to be at $kT \approx 5$~keV.

Given the two high quality spectra, we can compare the line fluxes directly and in a model-independent way. In order to do that, we need to account for the different line spread functions (LSF) of the instruments.  LETG has approximately a Gaussian LSF with $FWHM \approx 0.05 \AA$, while RGS has a slightly wider LSF of $FWHM \approx 0.07 \AA$, but a more complex shape with a narrow peak. For the comparison, we fold the fluxed LETG spectrum through the RGS response. For the details of this process see \citet{Nordon2005}.
In the following, we refer to the LETG spectrum, processed through the RGS response matrix, as the quiescent spectrum.

In Figure~\ref{fg:model_diff} we compare the flare and quiescent spectral lines directly.
We multiply the quiescent (LETG) spectrum by a factor of 1.25 to match the bright, isolated 19~\AA\ O$^{7+}$ Ly$\alpha$ line flux in the flare (RGS). Subsequently, a smooth phenomenological continuum is added to the rescaled quiescent spectrum to match the excess flare continuum. This enables us to compare only the {\it line} intensities. The flare and processed quiescent (i.e., rescaled + added continuum) spectra are plotted in Figure~\ref{fg:model_diff}. 
Also plotted is the residual of the two with the corresponding error bars.

\subsection{Emission measure distribution modeling}

%%%% Line Fluxes Table $$$$$$$$$$$
%\begin{longtable}{c c c c}
\begin{table*}
\caption{Measured line fluxes used for EMD fitting and abundance calculations.\label{tab:fluxtable}}
\begin{tabular}{c c c c c}
\hline \hline
Ion & Wavelength & Flare & Quiescence & Ratio \\
    & \AA & $10^{-4}$ photons cm$^{-2}$ s$^{-1}$ & $10^{-4}$ photons cm$^{-2}$ s$^{-1}$ & Flare/Quies.\\
\hline

Fe$^{16+}$ & 15.01   & 6.61 $\pm$  0.7     & 4.85  $\pm$ 0.50  & 1.36 $\pm$ 0.20 \\
Fe$^{17+}$ & 14.20   & 3.98 $\pm$  0.44    & 3.2   $\pm$ 0.33  & 1.25 $\pm$ 0.19 \\
Fe$^{18+}$ & 13.50   & 4.08 $\pm$  0.61    & 3.1   $\pm$ 0.32  & 1.32 $\pm$ 0.24 \\
Fe$^{19+}$ & 12.83   & 4.98 $\pm$  0.59    & 3.08  $\pm$ 0.32  & 1.62 $\pm$ 0.26 \\
Fe$^{20+}$ & 12.28   & 3.64 $\pm$  0.53    & 3.14  $\pm$ 0.33  & 1.16 $\pm$ 0.21 \\
Fe$^{21+}$ & 11.77   & 2.89 $\pm$  0.52    & 2.18  $\pm$ 0.23  & 1.33 $\pm$ 0.28 \\
Fe$^{22+}$ & 11.00   & 3.19 $\pm$  0.6     & 1.40  $\pm$ 0.15  & 2.29 $\pm$ 0.49 \\
Fe$^{23+}$ & 10.64   & 4.74 $\pm$  0.63    & 1.18  $\pm$ 0.13  & 4.02 $\pm$ 0.69 \\
Fe$^{24+}$ & 1.85    & 2.38 $\pm$  0.3*    & 0.136 $\pm$ 0.08  & 17.6 $\pm$ 10.5 \\
Fe$^{25+}$ & 1.78    & 0.51 $\pm$  0.16*   & ...               & ...             \\ 
N$^{6+}$   & 24.78   & 6.66 $\pm$  0.75**  & 3.54  $\pm$ 0.46**& 1.88 $\pm$ 0.32 \\
O$^{6+}$   & 21.60   & 2.37 $\pm$  0.37    & ...               & ...             \\
O$^{7+}$   & 18.97   & 21.8 $\pm$  2.2     & 17.4  $\pm$ 1.8** & 1.25 $\pm$ 0.18 \\
Ne$^{8+}$  & 13.45   & 3.55 $\pm$  0.75    & 3.53  $\pm$ 0.37  & 1.01 $\pm$ 0.24 \\
Ne$^{9+}$  & 12.13   & 22.1 $\pm$  2.3     & 16.0  $\pm$ 1.6   & 1.39 $\pm$ 0.20 \\
Mg$^{10+}$ & 9.17    & 0.94 $\pm$  0.27    & 0.88  $\pm$ 0.10  & 1.07 $\pm$ 0.33 \\
Mg$^{11+}$ & 8.42    & 4.82 $\pm$  0.66    & 1.56  $\pm$ 0.16  & 3.08 $\pm$ 0.53 \\
Si$^{12+}$ & 6.65    & 1.41 $\pm$  0.41    & 1.14  $\pm$ 0.12  & 1.24 $\pm$ 0.38 \\
Si$^{13+}$ & 6.18    & 1.37 $\pm$  0.7     & 1.12  $\pm$ 0.12  & 1.23 $\pm$ 0.64 \\
S$^{14+}$  & 5.05    & ...                 & 0.54  $\pm$ 0.07  & ...             \\
S$^{15+}$  & 4.73    & ...                 & 0.30  $\pm$ 0.05  & ...             \\
Ar$^{16+}$ & 3.96    & ...                 & 0.27  $\pm$ 0.04  & ...             \\
Ar$^{17+}$ & 3.73    & ...                 & 0.036 $\pm$ 0.03  & ...             \\

\hline
\end{tabular}
\begin{enumerate}
\item[*]{Extracted from EPIC-pn}
\item[**]{Used for abundances only}
\end{enumerate}

\begin{tabular}{p{11cm}}
{\it NOTE:} For He-like ions, the line flux includes the resonant line only. For H-like ions the flux includes both lines of the Ly$\alpha$ doublet. For Fe the flux includes all the transitions of the ion within 0.03\AA\ of the given wavelength. Flux errors include 10\%\ instrument calibration error. 
\end{tabular}

\end{table*}

The continuum is almost featureless for a distrubution of plasma temperatures and in addition, the continuum from the LETGS instrument is unreliable due to the CC mode. Therefore line fluxes alone are used to derive the emission measure distribution (EMD). Table~\ref{tab:fluxtable} lists the measurable lines in the flare and in quiescence.
The observed line flux $F^{q}_{ji}$ of ion $q$ due to the atomic transition $j \rightarrow i$ can be expressed by means of the element abundance with respect to Hydrogen $A_z$, the distance to the object $d$, the line power $P^q_{ji}$ and the ion fractional abundance $f_q$ as:

\begin{equation}
  F^q_{ji} = \frac{A_z}{4 \pi d^2} \int_{0}^{\infty}{P^q_{ji}(T) f_q(T) EMD(T) \mathrm{d}T }
  \label{eq:lineflux}
\end{equation}

We use the primary line (see Table~\ref{tab:fluxtable}) from every Fe ion in the observed spectra to get a set of integral equations (eq.~\ref{eq:lineflux}), whose solution yields the EMD scaled by the unknown Fe abundance. For other elements, we use ratios of the He-like to H-like line fluxes instead of absolute fluxes, thus the element abundance $A_z$ cancels out. This adds another set of equations that constrain the shape of the EMD, and do not depend on the abundances:

\begin{equation}
  R_z = \frac{ F^{He-like}_{ji} }{ F^{H-like}_{lk} }
  \label{eq:fluxratio}
\end{equation}

The X-ray spectra include as many as ten Fe ions but no more than two ions from other elements. In total, we get 14 equations for the flare and quiescence observations, but different equations, depending on which lines are visible (see Table~\ref{tab:fluxtable}). We fit the line fluxes and flux ratios using the {\it least squares best fit} method to solve for the EMD, where the EMD is expressed by a parameteric non-negative function of $T$. 
This method yields the estimated shape of the EMD, independent of any assumptions for the abundances, and is scaled by the Fe abundance. The integration in eq.~\ref{eq:lineflux} and eq.~\ref{eq:fluxratio} is cut-off at 8~keV beyond which the EMD is completely degenerate. This means that some of the EM in the last bin could be attributed to even higher temperatures.

The atomic data for the line powers are calculated using the HULLAC code \citep{HULLAC}. In order to measure the line fluxes and solve for possible blending, we preform an ion-by-ion fitting to the spectra. The line powers for each ion are calculated at its maximum emissivity temperature and then passed through the instrument response. The observed spectra are fitted by a set of complete individual-ion spectra simultaneously, resulting in an excellent fit that accounts for all the observed lines and blends. This process is similar to the one used in \citet{Behar2001} and \citet{Brinkman2001}.
The line fluxes used in the EMD fitting are listed in Table~\ref{tab:fluxtable}.
The ionic abundances ($f_q$) for: Fe, Ar, S, Si, Mg are taken from \citet{Gu2003}, whereas \citet{Mazzotta1998} is used for the other elements. 

Our goal is to compare the EMD of the flare and quiescence states. It is important to note that the solution for the EMD is not unique as is the case with integral equations of this sort \citep{Craig1976}. On scales much smaller than the width of the ions emissivity curves, or in temperature regions where there are no emissivity peaks of any ion, there is no way of constraining the EMD. Therefore, in order to be able to compare the EMD solutions, the confidence intervals of the solution are as important as the actual values.
We choose to fit a {\it staircase} shaped function to allow for local confidence intervals estimates. The confidence intervals are calculated using the inverse $\chi^2$ distribution, meaning we search the parameter space for the $\chi^2$ contour that gives a deviation from the best fit that corresponds to the requested confidence level.

\begin{figure}
%  \begin{center}
    \vspace{0.8cm}
    \resizebox{\hsize}{!}{\includegraphics{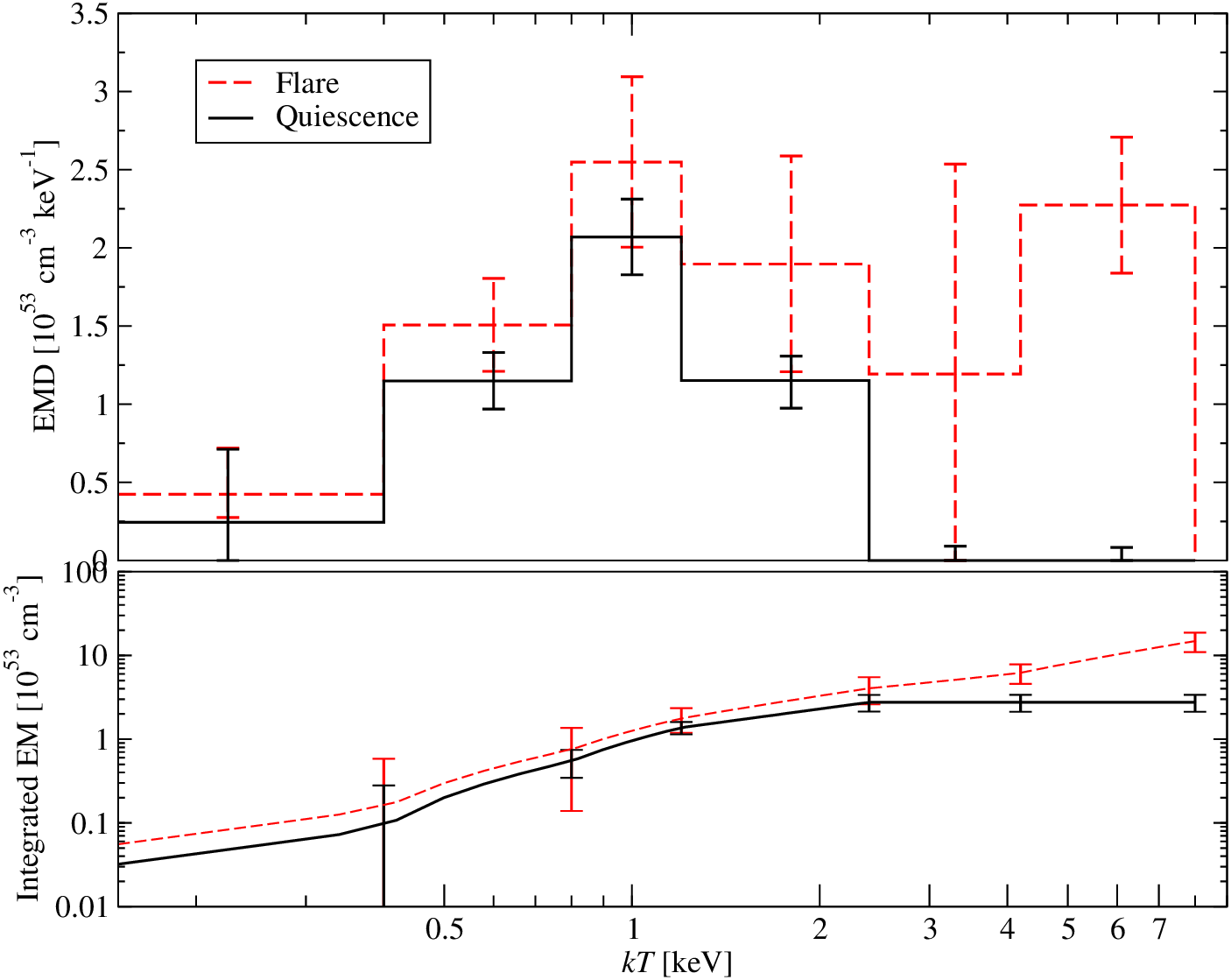}}
    \caption{{\it Top:} EMD of the two observations. Error bars indicate 90\%\ confidence intervals. {\it bottom:} The integrated EM up to $kT$, with 90\%\ confidence intervals. EMD is scaled according to Solar Fe abundance, taken to be: Fe/H = 4e-5.}
    \label{fg:EMD_plot}
%  \end{center}
\end{figure}

The fitted EMD is plotted in Figure~\ref{fg:EMD_plot} with 90\%\ confidence intervals. The value of the EMD in each bin represents the average EMD over the bin. Selecting the number of bins and their widths is not trivial. Since, as discussed above, we are interested in meaningful confidence intervals, we cannot use narrow bins, as this will result in excessive error bars. The line emissivity curves have considerable widths and some extend to temperatures much higher than their peak emissivity, resulting in strong negative correlations between the EM in neigbouring bins.
The errors on the measured fluxes increase the uncertainty on the EM even further. Ultimately, if meaningful confidence intervals are to be obtained, the number of bins has to be kept small and their width optimized according to the constraints in each region.

The important physical quantity is the {\it integral} of the EMD over a range of temperatures. 
The integrated EM from zero to $kT$ is plotted in the bottom panel of Figure~\ref{fg:EMD_plot} with 90\%\ confidence bars. The uncertainties caused by the strong correlation between the EMD bins disappear with integration, resulting in much smaller error bars.

In order to extract the X/Fe abundance ratios, we simply calculate the non-Fe line fluxes (eq.~\ref{eq:lineflux}) from the Fe-scaled EMD. The ratio between the measured and calculated flux gives the abundance value. The results for the elements observed by both instruments are summarized in Table~\ref{tab:table1}.
Errors for the abundances include both uncertainties in the EMD model and in the measured line fluxes.

%%%%%% Abundances table  %%%%%%%%%%
\begin{table}
\caption{The calculated abundances relative to Fe, obtained from the EMD analysis of the \cxc\ (quiescence) and \xmm\ (flare). ``Flare'' abundances are in fact averaged ``quiescence+flare'' components during the 2001 flare observed by \xmm. Uncertainties are 1$\sigma$. \label{tab:table1} }
\begin{tabular}{c c c c}
\hline \hline
        & Flare & Quiescence & Ratio \\
Element & X/Fe  & X/Fe & Flare/Quies. \\
\hline

N  & 15.1 $\pm$ 2.7  & 13.7 $\pm$ 2.3   & 1.1 $\pm$ 0.3 \\
O  & 30.3 $\pm$ 5    & 42   $\pm$ 6     & 0.72 $\pm$ 0.15 \\
Ne & 15.2 $\pm$ 2.3  & 19.9 $\pm$ 2     & 0.76 $\pm$ 0.14 \\
Mg & 1.94 $\pm$ 0.33 & 1.66  $\pm$ 0.17 & 1.17 $\pm$ 0.23 \\
Si & 0.99 $\pm$ 0.4  & 1.37  $\pm$ 0.14 & 0.72 $\pm$ 0.30 \\

\hline
\end{tabular}
\end{table}

%%%%%%%%%%%%%%%%%%%%%%%%%%%%%%%%%
\section{Discussion}
%%%%%%%%%%%%%%%%%%%%%%%%%%%%%%%%%

As seen in Figure~\ref{fg:model_diff}, the relative line intensities in flare and in quiescence match very well for wavelengths longer than 12~\AA. For the most part, the residual is less than 1$\sigma$. The fluctuations in the residual around the brightest lines at 19, at 13.5,  and at 12~\AA\ are caused by slight inaccuracies in the LSF of the RGS intensified by the strong steep peak. However, the residuals do average to zero. Around the 19~\AA\ line, that is by design of the scaling. 
Even at shorter wavelengths, the lines of the He-like Mg$^{10+}$ (9.17~\AA) and of Si$^{12+}$ (6.65~\AA), which have maximum emissivity temperatures of 540 and  860~eV, respectively, do not show detectable variations.

More evident changes in the spectrum start to appear below 12~\AA. The most conspicuous residuals are those due to lines of highly ionized iron: Fe$^{22+}$ and Fe$^{23+}$ between 10 and 12~\AA, whose emissivities peak at 1.2 and 1.6~keV, respectively. Since the relative intensities of the lower temperature Fe lines do not change during the flare, this is clearly a temperature effect rather than an abundance effect. Together with the dramatic change in the shape of the continuum, it indicates the appearance of a very hot component in the flare with little effect on the cooler plasma. This result is also seen from the EMD in Fig.~\ref{fg:EMD_plot}.

Another line that shows a significant change between flare and quiescence is the Ly$\alpha$ line of Mg$^{11+}$ at 8.42~\AA, whose flux increases by a factor of $2.4\pm0.4$ during the flare, beyond the overall 25\%\ enhancement. This change too is due to the high-T flare as explained in the next paragraph.
This comes to demonstrate that without clear identification of high-T lines, abundance effects can be easily confused with temperature changes, casting doubt on the reliability of some earlier results, which could not rely on line resolved spectra. 
The only other ion marginally in the RGS band that could probe temperatures above 1~keV is Si$^{13+}$, whose lines suffer from a low signal to noise ratio.

From the EMD comparison in Fig.~\ref{fg:EMD_plot}, we conclude that with high certainty, a large, hot component above 3~keV has risen during the flare and did not exist during quiescence. In the 2 - 4 keV region we get poor constraints on the flare EMD, due to the absence of lines with peak emissivity in this temperature region from the RGS band. This intensifies the negative correlation between bins mentioned above and results in a large error. 
For the quiescence EMD, we have Si, S and Ar lines that allow for much better constraints in this region.
At 2~keV and below, both EMDs seem to be similar, with the flare EMD being slightly higher. The quiescent LETG spectrum is missing lines from wavelengths longer than 20~\AA\ due to the aforementioned problems with CC mode. These lines would have allowed for much better constraints on the quiescence EMD below 1~keV.

We conclude that the plasma in its quiescent phase ($\leq$~2~keV) was not affected by the flare, except its emission measure increased by a constant factor of $\sim$1.25. We note that the large amount of high-T EM in the flare contributes some flux to lines of low-T H-like ions, since their emissivity extends to higher temperatures, much beyond the peak emissivity. This explains the added flux to the Mg$^{11+}$ (8.42~\AA) line. 
For the lines of the H-like ions O$^{7+}$ and Ne$^{8+}$, the flare component contributes $\sim 30\%$ of the total flux in the line. This is why our model requires that the O and Ne abundances decrease by that amount to maintain the relative quiescent flux level (Table~\ref{tab:table1}). 
No significant abundance variations beyond these are observed. 

This means that most likely, the 25\%\ brightening of the corona is due to gradual variations in the quiescent corona over the 15 months that passed between observations. In 11 observations by ROSAT over a similar time span, gradual changes of over 40\% in the total count rate were registered \citep{Yi}. These variations may correspond to the activity cycle of the system or changes in the active regions rather than to flares. \citet{audard01} have also reported that during a flare on HR~1099 the colder quiescent plasma was not affected. Both HR~1099 and $\sigma$ Gem are constantly very bright X-ray systems. Therefore, one expects to get a large emission contribution from the corona outside the flaring region.

Simple heating of the plasma does not change the total EM, therefore variations in the total EM indicate added material, changes in density or both. 
The total integrated EM up to 3~keV is slightly higher during the flare, although still consistent with the quiescence EM, within the 90\%\ confidence intervals.
The total EM up to the 8~keV cut-off during the flare is 5.4$\pm$1.9 times that of the quiescence EM, which is very large in a bright system such as \sg, so it is unlikely that such a huge amount of plasma is added to the corona. The more likely interpretation is that the hot EM originates from plasma heated lower in the chromosphere, where the higher density would result in a large EM, even for a small amount of evaporated plasma. 
The increase in density was not detected here, but this could be due to the high charge states (H-like and bare) typical of the high-T flare, for which no density diagnostics are available.

\citet{Sanz-Forcada2002} analysed the 1998 flare on \sg, detected by EUVE. 
The flare timescale of approximately 1 day, as well as the sharp rise and gradual decline are similar to those of the present flare.
They report an increase in line flux of highly ionized Fe by a factor of 2-4 and a continuum increase by a factor of 6. 
Their EM analysis shows that an excess EM is prevalent throughout a broad temperature range. However, the absence of error estimates for the EMD does not allow for unambiguous conclusions.
It is possible that the flare reported by Sanz-Forcada et al. was similar to the present one, only the EUV measurements were limited to probing temperatures of up to 1.5~keV, which is below the lower temperature end of the present flare. In the bottom panel of Fig.~\ref{fg:EMD_plot}, it can be seen that there is no appreciable effect on the integrated EM up to that temperature. 
The fact that no significant density increase was detected by \citet{Sanz-Forcada2002} may hint to the presence of much higher temperatures and ionization states in that flare, as were observed here.

%%%%%%%%%%%%%%%%%%%%%%
\section{Conclusions}
%%%%%%%%%%%%%%%%%%%%%%
There is little to no change in the chemical and thermal composition under 2~keV between the flaring and quiescent states. This seems to suggest that the flare did not have a significant influence on the coronal structure outside a limited flaring region. The effect of local flaring is similar to that which is observed on the sun. 
If the hot flaring component we observe is indeed, as suggested by the Neupert effect \citep{Gudel2002}, plasma evaporating from the chromosphere to the higher corona, then as it fills the coronal loop, the EM of the evaporated plasma will first increase in the loop as more material flows in, although it is already cooling, while in a later phase, it will cool back to chromospheric temperature and then condensate out of the corona, i.e., the EM will decrease again. During this process, most of the surrounding cooler plasma that we observe in the quiescent spectrum remains unaffected. As the flare decays, its contribution rapidly decreases and becomes insignificant to this quiescent emission.
The fact that the large flare had little influence on the cooler quiescent corona, suggests that the corona is stable against single large events and element enrichment processes are probably slow and continuous.

\acknowledgements
The research at the Technion was supported by ISF grant 28/03 and by a grant from the Asher Space Research Institute. PSI astronomy has been supported by the Swiss National Science Foundation (grant 20-66875.01). We thank Shai Kaspi for assistance with the data reduction and the referee Brian E. Wood for his useful comments. \xmm\ is an ESA science mission with instruments and contributions directly funded by ESA member states and the USA (NASA).

%%%%%%%%%%%%%%%%%%%%%%%%%%

\end{document}